\documentclass[12pt,a4paper]{article}
\usepackage[latin2]{inputenc}
\usepackage{amsmath}
\usepackage{physics}
\usepackage{amsfonts}
\usepackage{amssymb}
\usepackage{graphicx}
\usepackage{slashed}
\begin{document}

\title{\textbf{Emergent entropy of exotic oscillators and squeezing in three-wave mixing process}} 
\author{Sayan Kumar Pal\footnote{sayankpal@bose.res.in}$~^{1}$, 
	Partha Nandi\footnote{parthanandi@bose.res.in}$~^{1}$,
	Sibasish Ghosh\footnote{sibasish@imsc.res.in}$~^{2,3}$,\\
	Frederik G. Scholtz\footnote{fgs@sun.ac.za}$~^{4}$ and 
	Biswajit Chakraborty\footnote{biswajit@bose.res.in}$~^{1}$\\
{\small{$^{1}$S. N. Bose National Centre for Basic Sciences,}}\\
{\small{JD Block, Sector III, Salt Lake City, Kolkata 700106, India.}}\\
{\small{$^{2}$Optics \& Quantum Information Group, The Insititute of Mathematical Sciences,}}\\
{\small{C. I. T. Campus, Taramani, Chennai - 600113, India.}}\\
{\small{$^{3}$Homi Bhabha National Institute, Training School Complex,}}\\
{\small{Anushakti Nagar, Mumbai 400094, India.}}\\
{\small{$^{4}$Department of Physics,}}\\
{\small{University of Stellenbosch, Stellenbosch 7600, South Africa.}}}

\maketitle

\begin{abstract}
We demonstrate the existence of entanglement between the spatial degrees of freedom of a system of harmonic oscillators placed in the noncommutative Moyal plane (``exotic oscillators") by computing the entanglement entropy as measured by the von Neumann entropy of the reduced density matrix. It is explicitly verified that the entanglement arises from the noncommutativity, which controls the coupling strength between the spatial modes. This can easily be generalised to the case where the momentum components also satisfy noncommutative relations, so that the entire phase space becomes noncommutative. In the former case, i.e. when only the spatial noncommutativity is present, the underlying mathematical structure is reminiscent of the
Unruh effect, as observed by a Rindler observer whose acceleration now gets related to the noncommutative parameter. It is shown that the Landau problem in the presence of a harmonic interaction gives a concrete physical realisation of this effect. Finally, we show that phase-space noncommutativity can give rise to a the non-classical effect of squeezing, which results from the non-linearity of a medium in a three-wave mixing process.
\end{abstract}
 
\section{Introduction}

The concept of entropy and its universality in different physical systems has been a topic of general interest historically and more recently. Beautiful relations have transpired between quantum theories and gravitational physics through entropic considerations \cite{hawking}. The concept of entropy has now become universal and apart from thermodynamics and statistical mechanics, the closely related concept of entanglement entropy has begun to play a central role in diverse areas of physics from condensed matter systems to gravitation and quantum field theory \cite{sach, condense, cardy, witten, ashoksen}. Entropy is also a key concept in quantum information theory. In the seminal works of \cite{Koul, sred}, the entanglement entropy of coupled harmonic oscillators and free scalar fields were computed and related with black hole entropy. 

Closely related to these developments, deformed or quantum space-times have recently emerged as plausible candidates for the structure of space-time at the Planck scale \cite{dop, cam, sny}. In fact, the presence of space-time noncommutativity alone can have non-trivial consequences in nonrelativistic as well as relativistic quantum mechanics as has been shown very recently in \cite{firstann, verma}. Noncommutative spaces have also emerged in the low energy limit of certain string theories \cite{seiberg} and in the quantization of gravitating point particles in 2+1 dimensional space-time \cite{hooft}. Motivated by the above considerations, the present work aims to explore the role of entanglement entropy in deformed quantum systems such as the Moyal plane.  

Physics in (2+1) dimensions has been an exciting corner of research since there are many surprises that have unfolded over the years. In particular, the spin of fundamental particles in (2+1) dimensions can be any arbitrary real number. In fact, the system of identical quantum particles in (2+1)D cannot be exhaustively classified into just bosons and fermions. There can be other possible excitations whose statistics interpolates between bosons and fermions. Such particles are known as anyons and they follow exotic statistics because of their association to the Braid group, rather than the permutation group, and can also have arbitrary spin \cite{wilczek0}. These exotic particles have been found to be useful for the understanding of fractional quantum Hall effect where exotic statistics play an important role \cite{halperin}.

The recent experimental detection of anyons \cite{detect} in a particular heterogenous semiconductor has created a new surge of interest among theoreticians to venture more on these exotic particles. In fact, it has been well-known for a long time now that the Galilean group in 2+1 dimensions admits two central extensions \cite{levy}. In the seminal work \cite{jackiw}, these two central extensions were shown to be related with the mass and spin of the particle. As mentioned above, in 2+1 dimensions, spin needs not be integer or half-integer valued, it can be arbitrary \cite{jackiw, wilc} and one can anticipate the exotic physics of anyons to play a key role there. Indeed, in this scheme, the noncommutative parameter $\theta$ is identified with the second central extension of the planar non-relativistic Galilean algebra {\cite{duval, horvathy}}, which is basically a shadow of the relativistic spin in the nonrelativistic context. From a slightly different perspective,  the noncommutativity of position coordinates in 2+1 dimensions has been shown to be proportional to the spin parameter \cite{gerbert}. The noncommutative structure between coordinates of anyons basically helps us to understand the  fractional quantum Hall effect (FQHE),  where quantisation yields the wave functions proposed by Laughlin to describe the quantum states \cite{stichel}. The noncommutativity also helps us to compute the filling fractions at incompressibility in FQHE when each particle is attached with an infinitesimal magnetic flux tube corresponding to a singular magnetic field \cite{haldane100}.

The concept of entropy essentially goes back to the work of Rudolf Clausius, which was then formalised mathematically by Boltzmann in his fascinating eponymous Boltzmann formula and later vividly studied by Gibbs as the amount of randomness in the description of the microscopic states of a given definite system. Later von Neumann introduced the concept of entropy in quantum statistical mechanics through the use of density matrices which are operators on the Hilbert space of states \cite{von}. In the present paper, we work with von-Neumann's notion of entropy as applied to the reduced density matrix, also commonly referred to as the entanglement entropy, to demonstrate spatial entanglement of a two-dimensional harmonic oscillator in the deformed Moyal plane. Our formalism is based on the Fock space operator approach, unlike \cite{Koul, sred} where the entropy was calculated using wave-functions for the coupled 2D harmonic oscillators.  Recently the entropy of entanglement was also computed for coupled harmonic oscillators using a perturbative approach and polymer quantization \cite{subhajit}. In the current scenario, we find that this entropy is dependent on the noncommutative parameter $\theta$ and becomes zero as we move to the commutative, undeformed plane. The physical model that explicitly implements this setup is a very generic problem in condensed matter in which one considers the non-relativistic motion of a charged particle on a plane with a transverse applied external magnetic field and a harmonic potential.   We show the correspondence of this with the NC harmonic oscillator problem explicitly in subsection (3.2). Indeed in \cite{dipti}, the ground-state of many anyons in a harmonic potential in two dimensions has been computed, where it was found that the effect of exotic statistics between the particles takes the form of a long-range gauge interaction. By using the correspondence we arrive at an analog of the Unruh effect, as observed by a Rindler observer, say in (1+1)D quantum field theory, in our non-degenerate Landau problem. In particular, the Minkowski vacuum $|0>_M$, as observed by an inertial observer, now corresponds to the ground state $|0; \theta=0>$ of an isotropic oscillator placed in the commutative ($\theta=0$) plane and the vacuum state, as observed by the accelerated Rindler observer, now corresponds to the ground state $|0; \theta>$ of the oscillator placed in the NC Moyal plane, while the NC parameter $\theta$ itself gets formally related to the acceleration of the Rindler observer.

 Finally, we extend our results to the more general problem of noncommutative phase space where, in addition to noncommutative position coordinates, momentum components can also be noncommutative. This problem can be re-interpreted as the study of the well-known Landau problem in the presence of a harmonic interaction in the noncommutative Moyal plane. This is because the presence of noncommutative momenta mimics the presence of a magnetic field in the system since the gauge covariant derivatives can be interpreted as the noncommutative momenta in this setting. This physical realisation of the aforementioned system adds further motivation for the investigation carried out here, especially as  a study of the role of entanglement entropy in this physical system is lacking in the literature. As an additional motivation, we mention that noncommutative phase space is also one of the most natural approaches to combine quantum mechanics with general relativity \cite{Born, town}, following Born's principle of reciprocity. In fact, the presence of phase-space noncommutativity can lead to an increase in limiting mass of white dwarfs \cite{sayanB}. In the present context, as a byproduct, we also find that phase-space non-commutativity for the two 1D non-interacting harmonic oscillators gives rise to a two-mode squeezing effect (on the corresponding system of two 1D oscillators in the commutative phase-space), a well-known feature of non-classicality in the domain of Quantum Optics \cite{Scully}. Squeezing can be generated via a non-linear medium (e.g., parametric amplifier), the seed of such non-linearity can be qualitatively linked with the number of phase-space non-commutativity parameter(s).  The action of such a two-mode squeezing on the ground state of two oscillators in the commutative phase-space gives rise to the effect of enhancing the temperature of the individual oscillators and, as a result, an increase in entropy. It may be noted here that in \cite{Bernardini}, the issue of squeezing, generated by phase-space non-commutativity has been considered where the authors focused on the study of Wigner function on non-commutative phase-space  rather than connecting squeezing with phase-space non-commutativity through a non-linear process, as discussed here.
 
 The paper has been organised into two broad sections: a) Section 2 deals with position noncommutative space where, apart from presenting the computation of the entanglement entropy, we also discuss its similarity with the Unruh effect and b) Section 3 deals with a more general deformed structure, namely, phase-space noncommutativity and the entanglement entropy is again computed and shown to depend on a pair of noncommutative parameters. The connection with the squeezing phenomenon in quantum optics is also highlighted.

\section{Dynamics of exotic oscillators}
\subsection{Formulation on Noncommutative plane}

The phase space coordinates of a nonrelativistic particle moving in the noncommutative Moyal plane satisfy the following four-dimensional NC Heisenberg algebra, given by
\begin{eqnarray}
\label{NHA}
\left[\hat{x}_{i},\hat{x}_{j}\right]=i\theta_{ij} = i\theta \epsilon_{ij} ~ ; \quad \left[\hat{p}_{i},\hat{p}_{j}\right]=0~ ; \quad \left[\hat{x}_{i},\hat{p}_{j}\right] = i{\hbar}\delta_{ij}.
\end{eqnarray}
Here $\theta$ denotes the constant spatial noncommutative parameter that can be taken to be positive ($\theta>0$) without loss of generality. A consistent formulation of quantum mechanics with the above algebra was initiated in \cite{Sir} using Hilbert-Schmidt operators and subsequently a proper interpretation of position measurement, in coherent state basis, was provided through a weak measurement \cite{fgs}. However in the present work we make use of a different approach, where we shall be working with an effective commutative theory.\\

To that end, first observe that the above algebra can be realized in terms of commutative position coordinates $q_i$ by the well-known Bopp transformations \cite{Moyal} given as,
\begin{equation}\label{cv2}
\hat{x}_i=q_i-\frac{\theta}{2\hbar}\epsilon_{ij}\pi_j~ ; ~~~ \hat{p}_i=\pi_i
\end{equation}
 where the associated phase space variables ($q_i, \pi_j$) now satisfy the standard ``commutative" Heisenberg algebra:
\begin{eqnarray}
\label{HA}
\left[q_{i},q_{j}\right]=0 ~ ; \quad \left[\pi_{i},\pi_{j}\right]=0~ ; \quad \left[q_{i},\pi_{j}\right] = i{\hbar}\delta_{ij}.
\end{eqnarray}
Note that we are using over-head hat to denote noncommutative phase space operators, in contrast to their commutative counterparts, which are denoted without the hat notation.\\

The Hamiltonian of the NC harmonic oscillator in the Moyal plane is given by-
\begin{equation}
H = {1\over 2m}(\hat{p}_1^2 + \hat{p}_2^2) + {1\over 2} m\omega^2 (\hat{x}_1^2 +\hat{x}_2^2) \label{L4}
\end{equation}
The Hamilton's equations of motion, following from  this Hamiltonian is obtained as,
\begin{equation}\label{eqn1}
\ddot{\hat{x}}_i-\frac{m\omega^2\theta}{\hbar} \epsilon_{ij}\dot{\hat{x}}_j + \omega^2 \hat{x}_i =0
\end{equation}
The effective commutative Hamiltonian is now obtained by making use of (2) in (4) to get -
\begin{eqnarray}\label{hami1}
H=\frac{1}{2m_R}(\pi_1^2 + \pi_2^2) + \frac{1}{2} m\omega^2(q_1^2 +q_2^2)- \frac{m\omega^2 \theta}{2\hbar}(q_1 \pi_2 - q_2 \pi_1) 
\end{eqnarray} 
Note that the footprints of noncommutativity i.e. $\theta$ dependence can now be found from the renormalized kinetic mass $m_R$ occurring as coefficients in the kinetic term and also in the popped up ``Zeeman-like" term, where $m_R$ is given as,
\begin{equation}
\frac{1}{m_R}=\frac{1}{m}+\frac{m\omega^2\theta^2}{4\hbar^2}=\frac{\kappa^2}{m}~~;~\kappa=\sqrt{1+\frac{m^2 \omega^2 \theta^2}{4\hbar^2}} \label{kappa}
\end{equation}

To get the normal modes of the aforementioned NC system, we can now work with this effective commutative Hamiltonian and introduce the following sets of $\theta-$independent creation and annihilation operators -
\begin{equation}\label{commop}
	b_{i}=\sqrt{\frac{m\omega}{2\hbar}}(q_{i}+i\frac{1}{m\omega}\pi_{i}); ~i=1,2
\end{equation}
satisfying the commutation relation $[b_{i},b^{\dagger}_{j}]=\delta_{ij}$. In the commutative ($\theta\rightarrow 0$) limit, the Hamiltonian (\ref{hami1}) will correspond to a planar isotropic oscillator and can be diagonalized by these creation and annihilation operators ($b_i$ and $b_i^{\dagger}$). However, for the present case (where $\theta \ne 0$), the Hamiltonian in (\ref{hami1}) can only be diagonalized by suitably deformed creation and annihilation operators $(a_+(\theta), a_-(\theta))$ to be described below satisfying isomorphic commutation relations. In fact, at an intermediate stage, we can introduce such types of deformed operators to re-write the Hamiltonian (\ref{hami1}) as,
\begin{eqnarray}
	H=\hbar \omega\kappa \bigg({a}_{i}^{\dagger}(\theta) a_i(\theta) + 1\bigg)+ i\hbar \gamma \bigg(a_1^{\dagger}(\theta) a_2(\theta) - a_1(\theta) {a}_2^{\dagger}(\theta)\bigg);~\gamma=\frac{m\omega^2 \theta}{2\hbar}
	\label{ham1_0}
\end{eqnarray}
Here we have performed a Bogolyubov type canonical transformation i.e. a unitary
transformation \emph{a la} von-Neumann, at the operator level as,
\begin{equation}\label{bogonew}
	a_{i}(\theta) =U b_{i}U^{\dagger}=\cosh\phi ~b_i + \sinh\phi ~b_i^{\dagger},~ with ~\phi=\log(\sqrt{\kappa})
\end{equation}
where $ U= e^{\frac{\phi}{2}(b^{2}_{j}-b^{\dagger2}_{j})}$ , with
$[a_{i}(\theta),a^{\dagger}_{j}(\theta)]=\delta_{ij}.$
This transformation generates non-classicality in the individual modes: the single-mode squeezing transformation $ U$ with real squeezing parameter $\phi$.

At the final stage, we introduce the following pair of annihilation operators -
\begin{equation}
	\begin{pmatrix}
		a_1(\theta) \\
		a_2(\theta)
	\end{pmatrix}
	\longrightarrow
	\begin{pmatrix}
		a_+(\theta) \\
		a_-(\theta)
	\end{pmatrix}
	={\bf M}\begin{pmatrix}
		a_{1}(\theta) \\
		a_{2}(\theta)
	\end{pmatrix} = \frac{1}{\sqrt{2}}\begin{pmatrix}
		1  & -i \\
		i  & -1 \
	\end{pmatrix}
	\begin{pmatrix}
		a_{1}(\theta) \\
		a_{2}(\theta)
	\end{pmatrix} \label{u(2)}
\end{equation}
where $M\in U(2)$. These operators are related to operators in (\ref{commop}) using (\ref{bogonew}) as,
\begin{equation}\label{normalmode100}
	a_{+}(\theta)=\frac{1}{\sqrt{2}}(a_{1}(\theta)-ia_{2}(\theta))=U b_{+}U^{\dagger};~~~~ a_{-}(\theta)=i\frac{1}{\sqrt{2}}(a_{1}(\theta)+ia_{2}(\theta))=iU b_{-}U^{\dagger}   
\end{equation}
where $b_{+}=\frac{1}{\sqrt{2}}(b_{1}-ib_{2})$, $b_{-}=\frac{1}{\sqrt{2}}(b_{1}+ib_{2}).$ 
These new Fock space operators also satisfy -
\begin{equation}
	[a_{\pm}(\theta) , a_{\pm}^{\dagger}(\theta)]=1~ , [a_+(\theta) , a_{-}(\theta)]=0=[a_+^{\dagger}(\theta) , a_{-}^{\dagger}(\theta)]
\end{equation}
On working through the Hamiltonian (\ref{ham1_0}), one can see that it now takes the diagonal form :
\begin{eqnarray}
	H=\hbar \omega_+ \bigg({a}_{+}^{\dagger}(\theta) a_+(\theta) + \frac{1}{2}\bigg) + \hbar \omega_- \bigg({a}_{-}^{\dagger}(\theta) a_-(\theta) + \frac{1}{2}\bigg);~\omega_{\pm}=\omega\kappa \mp \gamma
	\label{ham1_f}
\end{eqnarray}
This represents a system of decoupled oscillators with underlying commutative phase-space where we can easily identify $\omega_{\pm}$ as the characteristic frequencies.
  
The ground state of the system can therefore be readily expressed as the tensor product of individual ground states as-
\begin{equation}\label{vacuumnc}
|0,0 ; \theta >=~ |0;\theta>_+ \otimes~ |0;\theta>_-~~;~a_{\pm}(\theta)|0;\theta^{}>_{\pm}=0
\end{equation}
and the non-degenerate energy eigen-states are given by
\begin{equation}
	\begin{aligned}
		&~~~~~~~~~~~~\left|n_{+},n_{-};\theta\right\rangle
		=\frac{\left({a}^{\dagger}_{+}(\theta)\right)^{n_{+}}\left({a}^{\dagger}_{-}(\theta)\right)^{n_{-}}}{\sqrt{n_{+}!}\sqrt{n_{-}!}}\left|0,0;\theta\right\rangle
		\label{sp}
	\end{aligned}
\end{equation}
It then follows from (\ref{vacuumnc}) that the ground state also satisfies the following relation -
\begin{equation}
{a}_{\pm}(\theta)\left|0,0;\theta\right\rangle =0\implies {b}_{\pm}(U^{\dagger}\left|0,0;\theta\right\rangle) =0.
\end{equation}
And therefore we have,
\begin{equation}\label{sqvac}
	\left|0,0;\theta\right\rangle= U\left|0,0;\theta=0\right\rangle ,
\end{equation}
\begin{equation}
	\rho(\theta)=\left|0,0;\theta\right\rangle\left\langle 0,0;\theta\right|=U(\left|0,0;\theta=0\right\rangle\left\langle 0,0;\theta=0\right|)U^{\dagger}=U\rho(\theta=0)U^{\dagger}
\end{equation}
which shows how the vacuum and vacuum state density matrix of the 2D noncommutative harmonic oscillator (exotic oscillator) is related to that of ordinary 2D commutative harmonic oscillator.
Now moving further, accordingly the eigenbasis corresponding to the commutative number operator $(b^{\dagger}_{+}b_{+}+b^{\dagger}_{-}b_{-})$ can be defined as a tensor product decomposition into commutative subsystems, $\cal{H}=\cal{H}_+ \otimes \cal{H}_-$ :
\begin{equation}
\mathcal{H}=\textrm{span}\{ |n_+;\theta=0\rangle_+\otimes|n_-;\theta=0\rangle_-= \frac{\left({b}^{\dagger}_{+}\right)^{n_{+}}\left({b}^{\dagger}_{-}\right)^{n_{-}}}{\sqrt{n_{+}!}\sqrt{n_{-}!}}\left|0,0;\theta=0\right\rangle\}_{n_+,n_-=0}^{\infty}
\end{equation}
Now, on using completeness relation of the state vectors $\left|n_{+},n_{-};\theta=0\right\rangle := |n_+;\theta=0\rangle_+\otimes|n_-;\theta=0\rangle_- \in \cal{H}$ in (\ref{sqvac}), we have the ground state of the original Hamiltonian (\ref{hami1}) as,

\begin{eqnarray}\label{expansion1}
	\left|0,0;\theta\right\rangle&=&\sum_{n_{\pm}=0}^{\infty} \left|n_{+},n_{-};\theta=0\right\rangle\left\langle n_{+},n_{-};\theta=0\right|U \left|0,0;\theta=0\right\rangle \nonumber\\
&=&\sum_{n_{\pm}=0}^{\infty} d_{n_{+},n_{-}}(\theta)~|n_+; \theta=0>_+ \otimes~ |n_-; \theta=0>_-
\end{eqnarray}
where the quantities $d_{n_{+},n_{-}}(\theta)$ are called Wigner d- matrix elements,
\begin{equation}
	d_{n_{+},n_{-}}(\theta)=\left\langle n_{+},n_{-};\theta=0\right|U \left|0,0;~\theta=0\right\rangle    
\end{equation}
These coefficients $d_{n_+,n_-}(\theta)$ can now be determined in a straightforward manner by imposing 
 $a_{\pm}(\theta) |0, 0; \theta> = 0$, for which it will be convenient to re-express $U$ in terms of $b_{\pm}$ (for details see equation (\ref{two-modesqu}) in Appendix) to find:

\begin{equation}
d_{n_+,n_-}(\theta)= (-1)^{n_+} \sigma^{n_+} \delta_{n_+,n_-} ~;~~\sigma=\tanh \phi
\end{equation} 

Finally substituting this $d_{n_+,n_-}(\theta)$ in (\ref{expansion1}), we arrive at the desired form of the normalized ground state as -
\begin{equation}\label{vacuum}
|0, 0; \theta>= \sqrt{1-\sigma^2}\sum_{n=0}^{\infty} ~ (-1)^n \sigma^n {|n; \theta=0}>_+ \otimes~ {|n; \theta=0}>_-
\end{equation}
 The corresponding ground state density matrix is then given by:
\footnotesize
\begin{eqnarray}
\rho&=&|0, 0; \theta><0, 0; \theta|  \nonumber \\
&=&(1-\sigma^2) \sum_{n,m=0}^{\infty}  (-\sigma)^{n+m} \bigg(|n; \theta=0>_+ \otimes~ |n; \theta=0>_-\bigg)\bigg(~_+<m; \theta=0| \otimes~_- <m; \theta=0|\bigg) \nonumber
\end{eqnarray}
\begin{eqnarray}\label{totalrho}
=(1-\sigma^2) \sum_{n,m=0}^{\infty} ~  (-\sigma)^{n+m}~ |n; \theta=0>_{+}<m; \theta=0| \otimes~ |n; \theta=0>_{-}<m; \theta=0|
\end{eqnarray}
\normalsize
These are the desired expressions of the ground state $|0, 0; \theta>$ and the corresponding density matrix in terms of states of the commutative planar oscillator. This is still a pure state and the associated entropy vanishes, since after all purity of a state is basis-independent. But what we intend to show here is that these two 1D harmonic oscillators are entangled with each other and the corresponding entanglement entropy is induced by the Zeeman-like term in (6), which in turn is induced by noncommutativity. We now focus our attention to sub-system `$+$' and carry out partial tracing over sub-system `$-$' to obtain the
reduced density matrix for the subsystem `+' as:
\begin{equation}\label{reddensity}
\rho_+  \equiv \rho_{r} = Tr_- \rho=(1-\sigma^2) \sum_{n=0}^{\infty} ~  (\sigma^2)^{n}~ |n; \theta=0>_{+}<n; \theta=0|
\end{equation}
which is clearly a mixed state density matrix as $\rho_r^2\neq \rho_r$ unlike the initial density matrix $\rho$ in (\ref{totalrho}). In this form, one can easily read-off the eigenvalues $p_n$ of this reduced density matrix $\rho_r$ as
\begin{equation}
 p_n=(1-\sigma^2) \sigma^{2n}
\end{equation}

Therefore, the von-Neumann entropy of this subsystem is then given by
\begin{eqnarray}
S=&&-Tr(\rho_{r} \log\rho_{r}) \nonumber\\
=&& -\sum_{n=0}^{\infty} p_n ~\log p_n \nonumber\\
=&& -\log(1-\sigma^2) - \frac{\sigma^2\log\sigma^2}{1-\sigma^2} \label{spatialentropy}
\end{eqnarray}
where,
\begin{equation}
\sigma=\tanh{\phi}=\frac{1-\kappa}{1+\kappa}=\frac{1-\sqrt{1+\frac{m^2 \omega^2 \theta^2}{4\hbar^2}}}{1+\sqrt{1+\frac{m^2 \omega^2 \theta^2}{4\hbar^2}}}
\end{equation}
\\

As a simple verification one can easily check that in the commutative limit ($\theta \rightarrow0$), the parameters $\kappa\rightarrow 1$ and $\sigma\rightarrow 0$. Consequently, the entropy also vanishes. So in this system, the non-zero entropy basically stems from noncommutativity between the two spatial degrees of freedom of the system and the physical meaning of the entropy expression is the following: If we were to somehow probe the NC oscillator system by measurement of only the commutative observables, we will encounter a noise signifying the existence of this non-zero entropy (\ref{spatialentropy}). Note that (\ref{u(2)}) is an U(2) transformation (i.e. a beam splitter like transformation) from the set of two input modes ${a_1(\theta), a_2(\theta)}$ to the set of normal modes ${a_+(\theta), a_-(\theta)}$ - both sets being considered in the background commutative plane satisfying standard Heisenberg algebra (\ref{HA}). Now the non-classical nature of the input states results from (\ref{bogonew}) and hence as a result of the transformation in (\ref{normalmode100}), the two-mode output state produced is nothing but the tensor product of two single-mode squeezed vacuum states, which will, in turn, become an entangled two-mode state. Thus we see that the transformation from the set of operators ${b_+, b_-}$ to the set ${a_+(\theta), a_-(\theta)}$ via (\ref{bogonew}) maps the two-mode vacuum state $|0, 0; \theta = 0>$ to a two-mode entangled state (with any given value of $\theta)$ in the commutative plane. Note that the seed of the aforesaid entanglement lies in the existence of any non-zero value of the non-commutativity parameter $\theta$. The state on the RHS of (\ref{sqvac}), which is nothing but the two-mode squeezed vacuum state, where the two-mode squeezing transformation corresponds to the transformations (\ref{normalmode100}) together. Thus we see that $|0, 0; \theta>$ has finally become a two-mode entangled state with respect to the division + and -  (i.e., division with respect to the mode operators $b_+$ and $b_-$). Note that the state on the RHS of (\ref{vacuum}) is entangled if and only if $\theta$ is non-zero. 

It becomes imperative to mention here that the reduced density matrix  (\ref{reddensity}) can be cast in a form which is manifestly in the form of a thermal density matrix, describing an ensemble of one dimensional harmonic oscillators. To extract an effective temperature, let us recall the form of a thermal mixed state density matrix for this one dimensional harmonic oscillator system with frequency $\omega$, which is given by,

\begin{equation}
\rho_{th}=\sum p_{n}~ |n;\theta=0><n;\theta=0|
\end{equation}
where
$p_{n}=(1-e^{-\frac{\hbar\omega}{k_BT}}) e^{-\frac{n\hbar\omega}{k_BT}}$

Now on comparing this with (\ref{reddensity}), we can readily identify the effective temperature T as-
\begin{equation}\label{efft}
\frac{1}{T}= -\frac{2k_B}{\hbar \omega}\log{|\sigma|}
\end{equation}
Furthermore, we also note that the mean value of the occupation number of the commutative 1D oscillator in this state $\rho_r$ (\ref{reddensity}) is

\begin{equation}\label{be1}
	<b_+^{\dagger} b_+>_{\rho_r}= \frac{1}{e^{\frac{\hbar \omega}{k_B T}}-1}
\end{equation}
which precisely corresponds to the expression occurring in Bose-Einstein distribution. This gives the energy distribution (in units of $\hbar\omega$) of the sub-system `+' with frequency $\omega$ in the state $\rho_r$. At this stage we observe that both the temperature $T$ (\ref{efft}) and the mean occupation number of the commutative oscillator $<b_+^{\dagger} b_+>_{\rho_r}$ (\ref{be1}) vanishes in the commutative limit, as it should be since the reduced density matrix is then a pure state.
Furthermore, it is intriguing to mention here that the whole mathematical framework presented so far is reminiscent of the Unruh effect in 1+1 D quantum field theory (see \cite{cris} for a review). Indeed the vacuum Minkowski state $|0>_M$ in a discretized version takes the form:
\begin{equation}
	|0>_M=\prod_{i=1}^{\infty} \sqrt{1-e^{\frac{-2\pi \omega_i}{a}}}\sum_{n_i=0}^{\infty} ~  e^{\frac{-n_i\pi \omega_i}{a}}~ |n_i; R>\otimes |n_i; L>
\end{equation}
where $a$ is the acceleration of the Rindler observer and ${\omega_i}$ being the set of discrete frequencies.
In fact by formally replacing ($-\sigma$) by $e^{-\frac{\pi \omega_i}{a}}$ in (19) in our single particle system gives the corresponding expression in (28) for the i-th mode and the effective temperature in the Rindler problem can also be obtained by this same replacement in (26), where one just needs to restore $\hbar$ and Boltzmann constant $k_B$ by dimensional consideration. All this stem from the very similar structure of the reduced density matrix for the right wedge, obtained by taking partial trace over the left wedge and is given by:
\begin{equation}
\rho_R=\prod_{i=1}^{\infty} \bigg((1-e^{\frac{-2\pi \omega_i}{a}})\sum_{n_i=0}^{\infty} ~  e^{\frac{-2\pi n_i \omega_i}{a}}~ |n_i; R> <n_i; R|\bigg)
\end{equation}
when compared with (\ref{reddensity}). We should point out in this context that we could have used other bases to expand the pure state in (24), instead of harmonic oscillator basis $|n_{\pm};\theta=0>$ corresponding to commutative subsystems $+$, $-$. In that case the resultant reduced density matrix will generically be an improper mixed state, as the corresponding mixedness may not arise from any statistical uncertainty. 
On the other hand our choice of division of the Hilbert space into $+,-$ modes is the most natural choice as in this case we can attribute the mixedness to the statistical uncertainty since the reduced density matrix can indeed be cast in the form of a thermal density matrix (29), thereby enabling us to associate an effective temperature (30). Importantly the Bogolyubov transformation plays a vital role here and this feature is the main common link between our analysis with that of Unruh effect problem, which could only be uncovered by making use of our choice of basis. The only shortcoming that remains is that the Minkowski vacuum $|0>_M$ in Rindler problem now gets mapped to our noncommutative vacuum $|0, 0; \theta>$ (\ref{vacuum}); ideally it should have been other way around. 

A criticism that may be raised against the above arguments is that the commutative coordinates may not be physical, but mathematical constructs of the noncommutative oscillator problem originally in noncommuting position coordinates and as such are not detectable with some physical detector.  
	In the next subsection we show that this situation can in fact be physically realized through a slight variant of the well-known Landau problem where the raising/lowering operators will now be parametrized by the applied external magnetic field $B$, rather than any noncommutative parameter by making use of a correspondence with the noncommutative oscillator problem in the Moyal plane.

\subsection{Commutative oscillator potential in presence of a magnetic field}

Firstly, we will now establish the correspondence between the Landau problem in the presence of harmonic potential and the two-dimensional NC oscillator system studied in previous subsection 2.1.
The Hamiltonian (\ref{hami1}) of the latter system can be unitarily transformed into the following form :
\small
\begin{eqnarray}\label{unitaryH}
H\rightarrow	H'=\tilde{U}H\tilde{U}^{\dagger}=\frac{1}{2m}(\pi_1^2 + \pi_2^2) + \frac{1}{2} m\omega^2\kappa^2(q_1^2 +q_2^2)- \frac{m\omega^2\theta}{2\hbar}(q_1 \pi_2 - q_2 \pi_1) 
\end{eqnarray}
\normalsize
where the unitary operator $\tilde{U}\in SU(1,1)$ and is given by $\tilde{U}=e^{-\frac{i\log(\kappa)}{2\hbar}(q_i\pi_i+\pi_i q_i)}$ \cite{partha}.
 As is well-known that the Hamiltonian of a charged particle moving on a plane with an external magnetic field $B$ acting normal to the plane and in presence of an additional external harmonic potential (with frequency $\omega$) takes the following form in the symmetric gauge ($A_i=-\frac{B}{2}\epsilon_{ij}q_j$) :
\begin{eqnarray}\label{landau1}
	H=\frac{1}{2m}(\pi_1^2 + \pi_2^2) + \frac{1}{2} m\Omega^2(q_1^2 +q_2^2)- \frac{\omega_c}{2}(q_1 \pi_2 - q_2 \pi_1) ~;~\omega_c=\frac{eB}{m}
\end{eqnarray}
where $\Omega=\omega \lambda$,~$\lambda=\sqrt{1+\frac{\omega_c^2}{4\omega^2}}$~. Note here $\pi_i$ is the canonical momentum satisfying the commutative Heisenberg algebra (\ref{HA}) and $e$ is the electric charge.\\

 
The Hamiltonians (\ref{unitaryH}) and (\ref{landau1}) are of similar form and become equivalent upon identifying:

\begin{equation}\label{map}
\theta=\frac{eB\hbar}{m^2\omega^2}
\end{equation}
Consequently the parameter $\lambda$ gets identified with $\kappa$ (\ref{kappa}).
In this scenario, the chargeless limit of the Landau problem implies the commutative limit of the Moyal plane, for any given constant value of magnetic field in the Landau problem with oscillatory interaction. This establishes the correspondence between Landau problem in the presence of harmonic potential and the two-dimensional NC oscillator system studied in previous subsection 2.1 and the corresponding chargeless/commutative limits. 
Now if the particle were neutral ($e=0$), it would not have felt the presence of the magnetic field. This will correspond to $\theta=0$ i.e. the commutative limit in the previous subsection. The Hamiltonian for this neutral particle will then be:
\begin{eqnarray}
	H=\frac{1}{2m}\vec{\pi}^2  + \frac{1}{2} m\omega^2 \vec{q}^2
\end{eqnarray}
which is an isotropic harmonic oscillator system in the plane for which appropriate Fock space operators are the usual 2D harmonic oscillator ladder operators ($b_i's$) given in (\ref{commop}). Given that we have established the mapping between the above two physical descriptions, we will henceforth work with the Hamiltonian (\ref{landau1}), which can be diagonalized just as in the previous subsection, by using the canonical Bogolyubov transformation -
\begin{equation}\label{landauoperator}
	a_{i}(B) =U b_{i}U^{\dagger}=\cosh\phi ~b_i - \sinh\phi ~b_i^{\dagger},~ with ~\phi=\log(\sqrt{\lambda})
\end{equation}
where $ U= e^{-\frac{\phi}{2}(b^{2}_{j}-b^{\dagger2}_{j})}$ , with
$[a_{i}(B),a^{\dagger}_{j}(B)]=\delta_{ij}.$
The diagonalized form of the Hamiltonian (\ref{landau1}) turns out to be:
\begin{eqnarray}
	H=\hbar \Omega_+ \bigg({a}_{+}^{\dagger}(B) a_+(B) + \frac{1}{2}\bigg) + \hbar \Omega_- \bigg({a}_{-}^{\dagger}(B) a_-(B) + \frac{1}{2}\bigg)
	\label{landaudiag}
\end{eqnarray}
where the characteristic frequencies $\Omega_{\pm}$ are given by, $\Omega_{\pm}=\Omega\pm \frac{1}{2}\omega_c$.
Here the normal mode operators $a_{\pm}(B)$ have been obtained by subjecting $a_i(B)$ a $U(2)$ transformation, just as in (\ref{normalmode100}) to get-
\begin{equation}
	\begin{pmatrix}
		a_1(B) \\
		a_2(B)
	\end{pmatrix}
	\longrightarrow
	\begin{pmatrix}
		a_+(B) \\
		a_-(B)
	\end{pmatrix}
	= \frac{1}{\sqrt{2}}\begin{pmatrix}
		1  & -i \\
		i  & -1 \
	\end{pmatrix}
	\begin{pmatrix}
		a_{1}(B) \\
		a_{2}(B)
	\end{pmatrix} 
\end{equation}
while from (\ref{landauoperator}), $a_{\pm}(B)$ are related to $b_i's$ in the following manner:
\begin{equation}
	a_+(B)=Ub_+U^{\dagger} ~;~a_-(B)=iUb_-U^{\dagger}
\end{equation} 
Inverting the above relations, we get-
\begin{equation}
	b_+=U^{\dagger}a_+(B)U=\cosh{\phi}~ a_+(B) + i\sinh{\phi}~   a_-^{\dagger}(B)
\end{equation}
\begin{equation}
	b_-=-i U^{\dagger}a_-(B)U=\sinh{\phi}~   a_+^{\dagger}(B)  -i\cosh{\phi}~   a_-(B)
\end{equation}

The ground state for the neutral particle is given by:
$b_{\pm}|0,0>_o=0$ from which it follows that-
\begin{equation}
a_{\pm}(B)(U|0,0>_o)=0 \implies |0, 0; B>=U|0,0>_o
\end{equation}
And therefore we have,
\begin{equation}
	|0,0>_o= U^{\dagger}|0, 0;B>
\end{equation}
Therefore one can write down the ground state $|0, 0>_o$ of the neutral particle as a linear combination of tensor products of `$\pm$' modes of the charged particle as :-
\begin{equation}\label{expansion}
	|0, 0>_o= \sqrt{1-\zeta^2}\sum_{n=0}^{\infty} ~ (-1)^n \zeta^n |n; B>_+ \otimes~ |n; B>_-
\end{equation}
where, $\zeta=\frac{\lambda-1}{\lambda+1}$. Note that we have used the subscript `o' to denote the ground state of a neutral particle. This expression (38) is the counterpart of (19) of the previous subsection.  Let us now consider the case where the applied magnetic field is very strong compared to the harmonic interaction i.e. $\omega_c>> \omega$. In this situation, we have:
\begin{equation}
	\Omega_+\simeq\omega_c+\frac{\omega^2}{\omega_c}~;~~ \Omega_-\simeq\frac{\omega^2}{\omega_c}
\end{equation}
obtained after keeping terms only upto first order in $\frac{\omega^2}{\omega_c^2}$. Since there is a large energy gap in the system, the higher states will be effectively inaccessible for the low energy particle and therefore we can trace out the higher energy modes $|n; B>_+$ to get the reduced density matrix as:
\begin{equation}
	\rho_{r} = Tr_+ \rho=(1-\zeta^2) \sum_{n=0}^{\infty} ~  (\zeta^2)^{n}~ |n; B>_{-}<n; B|
\end{equation}
which has the same form as (21) with $\sigma$ replaced by ($-\zeta$) ($\sigma \rightarrow (-\zeta$)). Therefore
if one follows the computations exactly as in the previous section and one can assign an expression of entropy-
\begin{eqnarray}
	S= -\log(1-\zeta^2) - \frac{\zeta^2\log\zeta^2}{1-\zeta^2}
\end{eqnarray}
and an effective temperature,
\begin{equation}
	\frac{1}{T}= -\frac{2k_B}{\hbar \omega}\log{|\zeta|}
\end{equation}
where we had to just replace $\sigma$ by ($-\zeta$) in (\ref{spatialentropy}) and (\ref{efft}) to get these expressions as the Fock vacuum of the neutral particle $|0>_o$ now appears as a mixed state, when expressed as a sum of tensor products of various Fock states of the charged particle.

Note that in contrast to the noncommutative oscillator, the commuting coordinates have the normal physical meaning here and an attempt by an observer to probe the system by measuring them would reveal the mixed state above.  To reveal the pure state that this mixed state derives from after the partial trace, the observer will have to make a `global' measurement that also probes the higher Landau levels, but these are inaccessible due to energy constraints. 

Owing to the correspondence that we have previously mentioned (\ref{map}), we can write:
\begin{equation}\label{vacuumcomm}
	|0, 0; \theta=0>=	|0, 0>_o= \sqrt{1-\sigma^2}\sum_{n=0}^{\infty} ~  \sigma^n |n; \theta>_+ \otimes~ |n; \theta>_-
\end{equation}
where the R.H.S. represents the states of an observer in the effective noncommutative world. 
The analogy of this situation with the physics of Unruh effect where an accelerated observer sees the Minkowski vacuum to be a thermal state is more transparent now. In fact, we have the noncommutativity playing the role of acceleration in the Rindler problem and $|0, 0; \theta=0>$ (\ref{vacuumcomm}) is playing the role of Minkowski vacuum. This is further illuminated by the fact that the mean occupation number $a_-^{\dagger}(B)a_-(B)$ in the state $\rho_r$ (42) of the neutral particle takes the form of Bose-Einstein distribution:
\begin{equation}
	<a_-^{\dagger}(B)a_-(B)>_{\rho_r}= \frac{1}{e^{\frac{\hbar \omega}{k_B T}}-1}
\end{equation}

In the next section, we will show how our present approach can be simply extended to show the emergence of entropy in a more generalized system where momentum components will also not commute apart from noncommuting position coordinates. This will also pave the way for relating it to the phenomenon of squeezing in a three-wave mixing process in a non-linear medium, as encountered in typical quantum optical scenario.


\section{Phase space noncommutativity}

In this section, we now consider a more general and natural class of noncommutative structures arising in phase space in which the associated operators follow the following NC Heisenberg algebra :
\begin{eqnarray}
\label{NHA}
\left[\hat{x}_{i},\hat{x}_{j}\right]=i\theta_{ij} = i\theta \epsilon_{ij} ~ ; \quad \left[\hat{p}_{i},\hat{p}_{j}\right]=i \eta_{ij}=i\eta \epsilon_{ij}~ ; \quad \left[\hat{x}_{i},\hat{p}_{j}\right] = i{\hbar}\delta_{ij}.
\end{eqnarray}
Here $\theta$ denotes the constant spatial noncommutative parameter that can be taken to be positive ($\theta>0$) without loss of generality and $\eta$ is the momentum noncommutative parameter, which should necessarily be intrinsically negative ($\theta\eta<0$) to ensure consistent quantization \cite{smy}. The phase space noncommutative structure is also thought to be a more consistent approach for unifying quantum mechanics and general relativity at high energy scales \cite{Born, town, Shahn}. In \cite{Shahn}, the occurrence of phase space noncommutativity at quantum gravity scales was rigorously argued using quantum groups and Hopf algebras. Furthermore, this kind of phase-space structure also arises in certain crystals like $GaAs$ exhibiting non-zero Berry curvature which leads to modified phase space density of states and thereby alters the transport properties \cite{niu, duvalberry}.

The algebra can be realized in terms of commutative position and momentum coordinates $q_i, \pi_i$ by the following generalized Bopp tansformations \cite{partha} given as -
\begin{equation}\label{cv2}
\hat{x}_i=q_i-\frac{\theta}{2\hbar}\epsilon_{ij}\pi_j + \lambda \epsilon_{ij} q_j~; ~~~ \hat{p}_i=\pi_i+\frac{\eta}{2\hbar}\epsilon_{ij}{q}_j + \lambda \epsilon_{ij} \pi_j
\end{equation}
where the associated variables now satisfy the standard Heisenberg algebra (\ref{HA}) with $\lambda=\frac{\sqrt{-\theta\eta}}{2\hbar}$. We now consider a harmonic oscillator in such a phase space whose Hamiltonian is also of the same form as that of (\ref{L4}). But now this can be re-written in terms of the commutative coordinates and momenta $q_i, \pi_i$  using (\ref{cv2}) as,
\begin{eqnarray}
H=\alpha(\frac{{\pi_1}^2 + {\pi_2}^2}{2m}) + {\beta}\frac{1}{2}m\omega^2(q_1^2 +q_2^2) +  \frac{\delta\omega}{2}(q_i \pi_i + \pi_i q_i)- \gamma(q_1 \pi_2 - q_2 \pi_1)
\label{hami11}
\end{eqnarray} 
where, 
\footnotesize
\begin{equation}
\alpha=1+\lambda^2 + \frac{m^2\omega^2\theta^2}{4\hbar^2}; ~~\beta=1+\lambda^2 + \frac{\eta^2}{4m^2 \omega^2\hbar^2};~ \gamma= \frac{\eta}{2m\hbar}+\frac{m\omega^2\theta}{2\hbar};\\
~~~~\delta=\lambda(\frac{\eta}{2m\omega\hbar}-\frac{m\omega\theta}{2\hbar})
\end{equation}
\normalsize
Let us recall (\ref{commop}) the creation and annihilation operators for the commutative oscillators of frequency $\omega$.
Solving the phase space operators $q$'s and $\pi$'s in terms of $b$ and $b^{\dagger}$, we have
\begin{eqnarray}
q_i=\sqrt{\frac{\hbar}{2m\omega}}\bigg( b_i+b^{\dagger}_i \bigg)~;~\pi_i=-i\sqrt{\frac{\hbar m\omega}{2}}\bigg( b_i-b^{\dagger}_i \bigg)
\end{eqnarray}
Now the Hamiltonian (\ref{hami11}) is rewritten using above to get-
\begin{eqnarray}
H=&&(\frac{2\beta}{\omega}+2\alpha\omega)(\frac{{b}_{i}^{\dagger}b_i+b_i{b}_{i}^{\dagger}}{4})+(\frac{\beta}{\omega}-\alpha\omega+2i\delta)\frac{b_i^{{\dagger2}}}{2}+(\frac{\beta}{\omega}-\alpha\omega-2i\delta)\frac{b_i^2}{2} \nonumber\\
&&+ i\hbar \gamma (b_1^{\dagger} b_2 - b_1 {b}_2^{\dagger}) \label{hami22}
\end{eqnarray}
Let us now invoke the following canonical transformation -
\begin{equation}
a_i(\theta, \eta)=\frac{1}{2}\bigg(\sqrt{\frac{\alpha}{\kappa'}}+\sqrt{\frac{\kappa'}{\alpha}}+i\frac{\delta}{\sqrt{\alpha\kappa'}}\bigg)b_i-\frac{1}{2}\bigg(\sqrt{\frac{\alpha}{\kappa'}}-\sqrt{\frac{\kappa'}{\alpha}}-i\frac{\delta}{\sqrt{\alpha\kappa'}}\bigg)b^{\dagger}_i
\end{equation}
with,
\begin{eqnarray}
 \kappa'=\sqrt{\alpha\beta-\delta^2}=\sqrt{1+2\lambda^2+\frac{m^2 \omega^2 \theta^2}{4\hbar^2}+\frac{\eta^2}{4m^2 \omega^2\hbar^2}}
 \end{eqnarray}
Note here that this transformation involves complex coefficients. However, by suitable parametrization this can be put in almost a similar form as in (\ref{bogonew}) -
\begin{equation}\label{bogo2}
a_i(\theta, \eta)=Ub_iU^{\dagger}=\cosh{\phi} e^{i\psi}b_i- \sinh{\phi} e^{-i\xi}b_i^{\dagger}
\end{equation}
where the unitary operator $ U= e^{-\frac{1}{2}(s b_j^2 -s^* b_j^{{\dagger2}})}$ and we have,
\begin{equation}
\tanh{\phi}=\sqrt{\frac{(\alpha-\kappa')^2+\delta^2}{(\alpha+\kappa')^2+\delta^2}} \nonumber
\end{equation}
\begin{equation}\label{mainpara}
\tan{\psi}=\frac{\delta}{\alpha+\kappa'} ~~;~~\tan{\xi}=\frac{\delta}{\alpha-\kappa'}~~,
\end{equation}
In terms of these new operators, the Hamiltonian $H$ (\ref{hami22}) turns out to be,
\begin{eqnarray}
H=\hbar \omega \kappa' \bigg({a}_{i}^{\dagger}(\theta, \eta) a_i(\theta, \eta) + 1 \bigg)+ i\hbar \gamma \bigg(a_1^{\dagger}(\theta, \eta) a_2(\theta, \eta) - a_1(\theta, \eta) {a}_2^{\dagger}(\theta, \eta)\bigg)
\label{ham2_0}
\end{eqnarray}
Finally making use of the $U(2)$ transformation introduced earlier in (\ref{normalmode100})-
\small
\begin{equation}\label{bogocross}
	a_+(\theta, \eta) =\frac{1}{\sqrt{2}}\bigg(a_1(\theta, \eta)-ia_2(\theta, \eta)\bigg)=Ub_+U^{\dagger}=\cosh{\phi} e^{i\psi}~ b_+- \sinh{\phi} e^{-i\xi}~ b_-^{\dagger}
	\end{equation}
\begin{equation}
	a_-(\theta, \eta)=\frac{i}{\sqrt{2}}\bigg(a_1(\theta, \eta)+ia_2(\theta, \eta)\bigg)=iUb_-U^{\dagger} =i\bigg[\cosh{\phi} e^{i\psi}~ b_- - \sinh{\phi} e^{-i\xi}~ b_+^{\dagger}\bigg] ,
\end{equation}
\normalsize
where $b_{\pm}$ has been defined earlier after (\ref{normalmode100}), thus facilitating the diagonalization of the Hamiltonian as -
\small
\begin{eqnarray}
H=\hbar \omega_+ \bigg({a}_{+}^{\dagger}(\theta, \eta) a_+(\theta, \eta) + \frac{1}{2}\bigg) + \hbar \omega_- \bigg({a}_{-}^{\dagger}(\theta, \eta) a_-(\theta, \eta) + \frac{1}{2}\bigg);\omega_{\pm}=\omega \kappa' \mp \gamma
\label{ham1_f}
\end{eqnarray}
\normalsize
Note that the spectrum of this system matches exactly with \cite{partha, fred}.
The ground state is now defined as-
\begin{equation}
|0, 0; \theta, \eta>~=|0; \theta, \eta>_+ \otimes~ |0; \theta, \eta>_-~~;~a_{\pm}(\theta, \eta)|0; \theta, \eta>_{\pm}=0
\end{equation}
It then follows from above that the ground state also satisfies the following relation -
\begin{equation}
	{a}_{\pm}(\theta, \eta)\left|0,0;\theta, \eta\right\rangle =0\implies {b}_{\pm}(U^{\dagger}\left|0,0;\theta,\eta\right\rangle) =0.
\end{equation}
And therefore we have,
\begin{equation}
	\left|0,0;\theta,\eta\right\rangle= U\left|0,0;\theta=0,\eta=0\right\rangle
\end{equation}
which relates the ground state of 2D harmonic oscillator in the NC phase space with that of 2D commutative harmonic oscillators.
Proceeding as in the previous section, the normalized ground state is finally written as-
\begin{equation}
	|0, 0; \theta, \eta>= \sqrt{1-\abs{\mu}^2}\sum_{n=0}^{\infty} ~ (-1)^n \mu^n {|n; \theta=0, \eta=0}>_+ \otimes~ {|n; \theta=0, \eta=0}>_-
\end{equation}
where $\mu$ is now a complex number and is given by,
\begin{equation} 
\mu=\tanh{\phi}~e^{-i(\psi+\xi)} \label{mainparameter}
\end{equation} 
where $\phi, \psi$ and $\xi$ have been previously defined in (\ref{mainpara}).

Again proceeding exactly in the same way as in previous section (2), the entropy associated with either one of the system is found to be-
\begin{eqnarray}
S=&&-Tr(\rho_{r} \log\rho_{r}) \nonumber\\
=&& -\log(1-\abs{\mu}^2) - \frac{\abs{\mu}^2\log\abs{\mu}^2}{1-\abs{\mu}^2} \label{generalentropy}
\end{eqnarray}\\
where $\mu$ has been defined in (\ref{mainparameter}). Equivalently, this expression is simply obtained by replacing $\sigma$ in (\ref{spatialentropy}) by $|\mu|$.

\subsection{Different limits of the entropy expression}
Firstly, on taking the momentum NC parameter $\eta$ to be zero, the function $\mu$ becomes real and equal to $\sigma$. So we recover (\ref{spatialentropy}) from the most general entropy expression (\ref{generalentropy}). On the contrary if we allow the spatial noncommutativity parameter $\theta$ to be zero, then we end up in a similar expression as (\ref{spatialentropy}) for the entropy in such a case but where $\kappa$ is now given by $\kappa=\sqrt{1+\frac{\eta^2}{4m^2 \omega^2\hbar^2}}$. \\

Furthermore, we see that the entropy vanishes on taking the limits $\theta$ and $\eta$ both going to zero. This is expected as then the system becomes that of a two-dimensional independent oscillators with no interaction and as a result there is zero  entropy.
This suggests that the expression (\ref{generalentropy}) is a very general one.
Moreover, the form of the entropy expression suggests that one can associate an effective non-zero temperature to the system. In such a case, the entropy expression (\ref{generalentropy}) becomes the thermal entropy of an ensemble of identical 1-D oscillators with frequency $\omega$ kept in contact with a thermal bath at temperature $\frac{1}{T}= -\frac{2k_B}{\hbar \omega}\log\abs{\mu}$, following similar arguments as in previous section (2). So in this sense noncommutativity can be interpreted to be the cause behind an emergent temperature.

\subsection{Realization of non-linearity in parametric amplification:} 
Here, we will discuss the role of squeezing transformations and how the set of calculations presented previously in this section might arise in the context of quantum optics \footnote{Needless to say that the scales of the noncommutative parameters $\theta$ and $\eta$ should now be adjusted accordingly and should clearly be many order of magnitudes above the one envisaged in, say, quantum gravity/Planck scale.}. We briefly describe here a process of relating the squeezing parameter (for single-and two-mode squeezing) with
non-linear properties of crystals, used for generating squeezing. Details of computations relating to squeezing transformations have been discussed in the Appendix. There are several ways for generating squeezing: using (i)parametric down conversion (three-wave or four-wave mixing), (ii)photon emissions by transions between pairs of
different energy levels of the atom, or (iii)optical fibres (three-wave mixing), etc. Non-linearity (that is, non-linear
dependence of polarization of light inside the crystal on the incoming electric field) is at the heart of each of these methods.


The Hamiltonian for non-degenerate parametric amplification in the interaction picture in the fully quantum-mechanical description is \cite{Scully}-
\begin{equation}
	H=\hbar g (a_s^\dagger a_i^\dagger b_p + a_s a_i b_p^\dagger)
\end{equation}
where g is the coupling constant, which is a function of second order non-linearity parameter $\chi^{(2)}$.

\begin{figure}[h]
	\setlength{\unitlength}{0.14in} 
	\centering 
	\begin{picture}(25,14)
\put(5,4){\framebox(12,6){\small{Non-linear crystal} ($\chi^{(2)}$)}}
\put(-3,6.75){\vector(2,0){8}}\put(17,8){\vector(1,0){8}}
\put(17,5.5){\vector(1,0){8}}
\put(-1,7) {$\nu_{pump}$ ($b_p$)}\put(19.5,4.5) {$\nu_{idler}$ ($a_i$)} \put(19,8.5)
{$\nu_{signal}$ ($a_s$)}
\end{picture}
	\caption{Three-wave mixing in parametric amplification} 
	\label{fig:lnlblock} 
\end{figure}
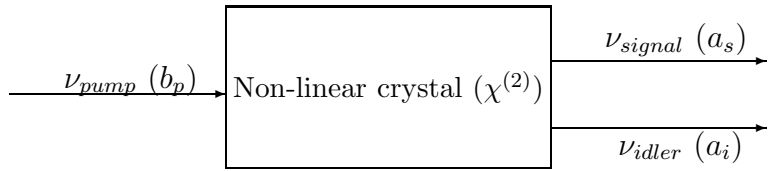


Now in parametric amplification one considers the pump field to be classical, so that we can replace $b_p$ by $\beta_p e^{i\phi_p}$ where $\beta_p$ is the amplitude and $\phi_p$ is the phase of coherent pump field. Under this situation, we have
\begin{equation}
	H=\hbar g \beta_p(a_s^\dagger a_i^\dagger e^{-i\phi_p} + a_s a_i e^{i\phi_p})
\end{equation}
This condition is valid when:
$gt\longrightarrow 0, \beta_p\longrightarrow \infty$ such that $g\beta_p t=$ constant, where $t$ is the time in the travelling wave reference frame.


		

	As a matter of fact, the non-linearity parameter $\chi^{(2)}$ of the crystal appears as an argument of the coupling parameter $g$ - in the case of squeezing (single as well as two-mode). An explicit form of g, in the case of two-mode squeezing via non-degenerate parametric amplification in the context of three wave mixing  (as depicted in Figure 1) is given by \cite{Scully, Lvovsky, coldatom} -
		\begin{equation}
			g\beta_pt=\chi^{(2)} \beta_p t= \chi^{(2)} |\Omega| \frac{nl}{c}
		\end{equation}
		where $|\Omega|\equiv \beta_p=$amplitude of input pump, $n=$refractive index of the crystal, $l=$length of the crystal, $c=$speed of light in vacuum. Thus, in this case of three-wave mixing, the associated squeezing parameter for two-mode squeezing, is given by:
		\begin{equation}
			|s|=\chi^{(2)} |\Omega| t=\chi^{(2)} |\Omega| \frac{nl}{c}
		\end{equation}
		
		So there is an explicit linear dependence of the squeezing amount $s$ on the crystal non-linearity parameter $\chi^{(2)}$. As here $\beta_p\neq0$ and $l\neq0$, therefore, $s=0$ iff $\chi^{(2)}=0$. Note that the parametric amplification process, depicted in Figure (1), is probabilistic in nature (with low success probability \footnote{It may be noted here that recently, there is a theoretical proposal for almost deterministic implementations of non-linear processes like three-wave mixing, four-wave mixing, etc. in the context of light-matter interactions. And such proposals have also the potentiality of being implemented in the lab with the presently available technology [``Deterministic quantum nonlinear optics with single atoms and virtual photons" by A. F. Kockum et al., Phys. Rev. A {\bf 95}, 063849 (2017)].
		}). Thus, the `effective' Hamiltonian $H$ in (68) (equivalently, (69)) works only when the parametric amplification process is a successful one, the latter occurs when certain conservation laws (energy conservation, momentum conservation, etc.) hold good \cite{Scully, Lvovsky, coldatom}. Again the Bogolyubov transformation (\ref{bogo2}) can be thought of as a squeezing transformation (\ref{squu}) [see Appendix] with a squeezing parameter $s$ which depend upon the noncommutativity parameters $\theta$ and $\eta$, and as a result of which the non-linearity parameter $\chi^{(2)}$ gets naturally related with the parameters $\theta$ and $\eta$ (this, as discussed above, holds as the two-mode squeezing happens due to the presence of the parameter ${\chi}^{(2)}$). \emph{Thus one may speculate of the structure of the phase-space inside the non-linear crystal to be of non-commutative nature in the case when the parametric amplification process in Figure (1) succeeds}. As mentioned above, although we considered above generation of squeezing only via parametric amplification (and thereby using non-linear crystal), all the other available resources for squeezing do require non-linearity. 
		It is to be mentioned in this context that in case of position non-commutativity alone i.e. if $\eta=0$ and $\theta\neq0$ as in previous section 3, we have a two-mode squeezing transformation with a real squeezing parameter $s=\frac{1}{4}\log(1+\frac{m^2 \omega^2 \theta^2}{4\hbar^2})$ as opposed to a complex parameter $s$ of the form, $s=\phi e^{i(\psi+\xi)}$ when both kinds of noncommutativities are present. Also there is an \emph{increase} in the squeezing amount if both $\theta\neq0$ and $\eta\neq0$ as compared with only position $(\theta\neq0,\eta=0)$ or momentum $(\eta\neq0,\theta=0)$ noncommutativity as the additional contributions are also positive-definite.
		

\section{Conclusion}
 Here in this paper, we have demonstrated the emergence of non-zero entropy in the ground state of the individual subsystems of a 2D oscillatory system when placed in deformed spaces, viz Moyal plane and noncommutative phase spaces. In particular, this means that if we focus on a particular mode of the exotic oscillator system and probe it irrespective of the state of the second system, then as the two modes are interacting with each other due to noncommutativity, tracing out one mode i.e oscillator leads to a loss of information or noise in the mode being probed. The probed mode behaves as if it is a system of commutative 1D oscillators interacting with an external heat bath at a specified temperature. In a noncommutative phase space as this, it is not possible to simultaneously probe the two oscillators with perfect precision and this fact manifests itself in the form of a non-zero entropy in our analysis. This provides a physical understanding of our present results. In sub-section 2.2, we have exploited the equivalence of Landau system with oscillatory interactions and the 2D noncommutative oscillator system to compute the entropy of neutral oscillators as perceived by a charged harmonic oscillator in presence of magnetic field. In other words, the charged particle observes the vacuum of the neutral particle as a thermal state. Here we have also shown how our observation enjoys an intriguing similarity with Unruh effect, where an accelerated Rindler observer finds the Minkowski vacuum state, as observed by an inertial observer, to be a thermal state and thereby relating the NC parameter to the acceleration of the Rindler observer. In either case, the corresponding thermal states can be assigned with effective temperatures and the expectation value of the occupation number operators have the form of Bose-Einstein distribution. It will be interesting to test such effects in cold Rydberg atom systems where position NC arises \cite{baxter}. 
 Note that our motivation for considering position noncommutativity is from the peculiar nature of spin parameter in (2+1) dimensions. Therefore for the general case of phase space NC, our results suggest that an anyonic charged particle sees the vaccum of the spinless chargeless particle as a mixed state.
 We have followed Fock space approach to compute the entropy. It will be an interesting task to investigate the system in the path integral scheme in the future. Recently, in \cite{polychro} the authors have computed entanglement entropy in noncommutative hyperplane and hypersphere. We plan to extend our analysis to such scenarios which have a similar spectrum as the spin Kitaev models. 
 We have also interpreted (and related qualitatively) the emergence of two-mode squeezing in the system of two oscillators as a result of the presence of phase-space non-commutativity in the non-linear medium, producing the squeezing effect. Probing the effects of deformation of quantum systems has been a challenging task and we hope to return to more of these issues in our future works.

\section*{Acknowledgments}
 S.K.P. and P.N. would like to extend their gratefulness to Department of Theoretical Physics, IMSc, Chennai for providing nice hospitality where the work was initiated. Also, the authors would like to thank Prof. A.P. Balachandran for his encouragement and for valuable discussions regarding this work. P. N. expresses his gratitude to Prof. Debnarayan Jana for an enlightening discussion on the subject. One of the authors, S.K.P. would like to thank UGC-India for providing financial support in the form of fellowship during
the course of this work.

\section*{Appendix}
Here, we are going to provide a brief description regarding the connection between single-mode and two-mode squeezings, well-known in the field of Quantum Optics.
Let us consider the relation between the single-mode non-commutative annihilation operators $a_i(\theta, \eta)$ and single-mode commutative annihilation operators $b_j$ (\ref{bogo2}). This transformation can be realized as a squeezing transformation with the corresponding single-mode squeezing operator being given by,\\
\begin{equation}\label{squeeze}
S^{(1)}_{b_j}(s)=e^{-\frac{1}{2}(s b_j^2 -s^* b_j^{{\dagger2}})}~~;~j=1,2.
\end{equation}
Here subscript 1 corresponds to single-mode squeezing. The complex squeezing parameter $s$ is given by 
\begin{equation}
	s=\phi e^{i(\psi+\xi)}~~; ~~~~\phi=\frac{1}{4}\log(1+2\lambda^2+\frac{m^2 \omega^2 \theta^2}{4\hbar^2}+\frac{\eta^2}{4m^2 \omega^2\hbar^2})
\end{equation}
  and $\psi, \xi$ as previously defined in (\ref{mainpara}). In fact, using the disentanglement theorem of SU(1, 1) \cite{Ban}, it can
be shown that-
\begin{equation}\label{squu}
a_j(\theta, \eta)=S^{(1)}_{b_j}(s) ~b_j ~(S^{(1)}_{b_j}(s))^{\dagger}
\end{equation}
Finally, the relation (\ref{bogocross}) between the normal-mode annihilation operators ($a_{\pm}(\theta, \eta)$) and commutative Fock operators can be realized by the following two-mode complex squeezing transformation:
\begin{equation}
\begin{pmatrix}
		a_+(\theta, \eta) \\
		a_-(\theta, \eta)
	\end{pmatrix}
	= \begin{pmatrix}
		e^{i\psi}\bigg[cosh(\phi)b_+ - sinh(\phi)	e^{-i(\psi+\xi)}b_-^{\dagger}\bigg] ~\\
		i	e^{i\psi}\bigg[cosh(\phi)b_- - sinh(\phi)e^{-i(\psi+\xi)}b_+^{\dagger}\bigg]~
	\end{pmatrix}
\end{equation}
In other words, this is a complex Bogolyubov transformation corresponding to the action of the two-mode squeezing operator $S^{(2)}_{(b_+, b_-)}(s)=e^{-(sb_+b_--s^*b_+^{\dagger}b_-^{\dagger})}$ i.e.
\begin{equation}\label{two-modesqu}
	\begin{pmatrix}
		a_+(\theta, \eta) \\
		a_-(\theta, \eta)
	\end{pmatrix}
	= S^{(2)}_{(b_+, b_-)}(s)\begin{pmatrix}
		b_+ \\
		ib_-
	\end{pmatrix}(S^{(2)}_{(b_+, b_-)}(s))^{\dagger}
\end{equation}
with the complex squeezing parameter given by $s=\phi e^{i(\psi+\xi)}$, which is the same as that in the single mode squeezing case. It can be shown easily that :

\begin{equation}\label{two-mode}
S^{(2)}_{(b_+, b_-)}(s)= U_{B'}\bigg(S^{(1)}_{b_+}(s) \otimes S^{(1)}_{b_-}(s) \bigg)U_{B}
\end{equation}
where $U_B'$ and $U_B$ are the unitary operators corresponding respectively to the lossless beam splitter matrices:
\begin{equation}B'=
\begin{pmatrix}
		\frac{1}{\sqrt{2}}  & \frac{1}{\sqrt{2}} \\
		\frac{i}{\sqrt{2}}  & -\frac{i}{\sqrt{2}} \
	\end{pmatrix}~;~~B=
	\begin{pmatrix}
		\frac{1}{\sqrt{2}}  & -\frac{i}{\sqrt{2}} \\
		\frac{i}{\sqrt{2}}  & -\frac{1}{\sqrt{2}}
	\end{pmatrix} 
\end{equation}
Thus we see from (\ref{two-mode}) that any two-mode squeezing transformation is related to a single-mode squeezing transformation -
with the same amount of squeezing -- via the actions of two fixed 50:50 lossless beam splitters. As lossless beam
splitters can neither create nor can it destroy non-classicality (in terms of photon number statistics), therefore, the actual
sources of the non-classical operation of squeezing here are the position and momentum observables non-commutativity parameters $\theta$ and $\eta$ respectively.
In fact, it follows directly from the aforesaid expression of the squeezing parameter that it's value is 0 if and only if $\theta, \eta=0$. The non-classicality in two-mode squeezing is actually generated by the non-classicality present in the single-mode squeezing with complex squeezing parameter $s=\phi e^{i(\psi+\xi)}$. Now, in case of only position noncommutativity $\eta=0, \theta\neq 0$, we also have a squeezing transformation as in (\ref{squu}) with the real squeezing parameter $s=\frac{1}{4}\log(1+\frac{m^2 \omega^2 \theta^2}{4\hbar^2})$ corresponding to (\ref{bogonew}) in section 2.1.\\

It may be recalled here that corresponding to the action of any lossless beam splitter with associated (unitary) beam splitter matrix,

\begin{equation}B(\Theta, \Phi, \Psi)= 
	\begin{pmatrix}
		\cos{\frac{\Theta}{2}} e^{i(\Psi+\Phi)/2}  & \sin{\frac{\Theta}{2}} e^{i(\Psi- \Phi)/2} \\
		-\sin{\frac{\Theta}{2}} e^{-i(\Psi- \Phi)/2}  & \cos{\frac{\Theta}{2}} e^{-i(\Psi+\Phi)/2}
	\end{pmatrix}~,
\end{equation}
its unitary representation $U_{B(\Theta, \Phi, \Psi)}$ is given by (with input mode operators $a_1, a_2$) :
\begin{equation}
U_{B(\Theta, \Phi, \Psi)}=e^{-i\Phi L_3} e^{-i\Theta L_2} e^{-i\Psi L_1}
\end{equation}
where, $L_1=\frac{1}{2} ({a}_1^\dagger a_2 + {a}_2^\dagger a_1); L_2=\frac{1}{2i} ({a}_1^\dagger a_2 - {a}_2^\dagger a_1); L_3=\frac{1}{2} ({a}_1^\dagger a_1 - {a}_2^\dagger a_2)$ are the three generators of su(2) algebra.


\begin{thebibliography}{73}
	\bibitem{hawking}
	J. D. Bekenstein, Phys. Rev. D 7, 2333 (1973); Phys. Rev. D 9, 3292 (1974); S. W.
	Hawking, Comm. Math. Phys. 43, 199 (1975)
	\bibitem{sach}S. Sachdev and J. Ye, Phys. Rev. Lett. \textbf{70(21),} 3339 (1993); A. Kitaev, A simple model of quantum holography, Talk given at the Fundamental
	Physics Prize Symposium (Nov 10, 2014); S. Sachdev, Phys. Rev. X \textbf{5,} 041025 (2015).
	\bibitem{condense}
	N. Iqbal, H. Liu, and M. Mezei, Lectures on Holographic
	Non-Fermi Liquids and Quantum Phase Transitions, in
	Proceedings of the 2010 Theoretical Advanced Study
	Institute in Elementary Particle Physics Boulder, Colorado,
	2010, edited by M. Dine, T. Banks, and S. Sachdev, World Scientific, Singapore, pp. 707$-$815 (2012).
\bibitem{cardy}P. Calabrese, J. Cardy, J. Stat. Mech. \textbf{0406} 06002 (2004).
	\bibitem{witten}Edward Witten, Rev. Mod. Phys. \textbf{90,} 045003 (2018).
	\bibitem{ashoksen}
	A. Sen,
	J. High Energy Phys \textbf{11} 075 (2008).

\bibitem{Koul}L. Bombelli, R. K. Koul, J. Lee, and R. Sorkin, Phys. Rev. D \textbf{34}, 373 (1986).
\bibitem{sred}M. Srednicki, Phys. Rev. Lett. \textbf{71,} 666 (1993).
\bibitem{dop}S. Doplicher, K. Fredenhagen, and J. E. Roberts, Phys. Lett. B \textbf{331,} 39 (1994).
\bibitem{cam}G. A. Camelia, Phys. Lett. B, \textbf{510,} 255 (2001).
\bibitem{sny}H.S Snyder, Phys. Rev. \textbf{71,} 38 (1947).
\bibitem{firstann}P. Nandi, S. K. Pal, A. N. Bose, B. Chakraborty, Ann. Phys. \textbf{386,} 305-326 (2017).
\bibitem{verma}R. Verma, P. Nandi, Gen. Rel. Grav. \textbf{51,} 143 (2019).
\bibitem{seiberg}N. Seiberg, E. Witten, J. High Energy Phys. 09 032 (1999).
	\bibitem{hooft}G. t'Hooft, Class. Quant. Grav. \textbf{13} 1023 (1996).
	
	\bibitem {rabin10}R. Banerjee, Mod. Phys. Lett. A \textbf{17,} 631-645 (2002).
\bibitem{wilczek0}F. Wilczek, Phys. Rev. Lett. \textbf{49,} 14 957 (1982).







\bibitem{halperin}B. I. Halperin, Phys. Rev. Lett. \textbf{52,} 1583 (1984); F. D. M. Haldane, Phys. Rev. Lett. \textbf{107,} 116801 (2011).
\bibitem{detect}H. Bartolomei \emph{et. al.} , Science \textbf{368,} 173-177 (2020); B. Rosenow, I. P. Levkivskyi, B. I. Halperin, Phys. Rev. Lett. 116, 156802 (2016).
	\bibitem{levy}J. M. Levy-Leblond, Group Theory and Applications, Academic Press, New
	York (1972); S. K. Bose, Comm. Math.
	Phys. \textbf{169}, 385 (1995).
	\bibitem{jackiw}R. Jackiw, V.P. Nair, Phys. Lett. B, \textbf{480,} 237 (2000).
	\bibitem{wilc}R. Mackenzie, F. Wilczek, Int. J. Mod. Phys. A, \emph{3,} 12 2827 (1988).
		\bibitem{duval}C. Duval, P. A. Horvathy,	\emph{J. Phys. A} \textbf{34}  10097 (2001).
			\bibitem{horvathy}P. A. Horvathy,	\emph{Ann. Phys.}, \textbf{299} 1  128 (2002).
			\bibitem{gerbert}P. S. Gerbert, Nucl. Phys. B, \textbf{346,} 440 (1990).
			\bibitem{stichel}P. A. Horvathy, L. Martina and P. C. Stichel,
			SIGMA {\bf 6,} 060  (2010).
			\bibitem{haldane100}F. G. Scholtz, B. Chakraorty, S. Gangopadhyay, J. Govaerts, \emph{J. Phys. A}, \textbf{38} 9849 (2005).
			\bibitem{von}J. von Neumann, Mathematical Foundations of Quantum Mechanics, Princeton University Press. (1955).
			\bibitem{subhajit}S. Barman, G. Sardar, 	Phys. Rev. D \textbf{99,} 125015 (2019).
			\bibitem{dipti}R. Chitra, D. Sen, 	Phys. Rev. B \textbf{46,} 17 10923 (1992).
		
	\bibitem{Born}M. Born, Proc. R. Soc. Lond. A \textbf{165,} 291-303 (1938).
	
	\bibitem{town}P. K. Townsend, Phys. Rev. D \textbf{15,} 10 2795 (1977).
	\bibitem{sayanB}S. K. Pal, P. Nandi, Phys. Lett. B \textbf{797,} 134859 (2019). 
		\bibitem{Scully}M. O. Scully, M. S. Zubairy, ``Quantum Optics", Cam. Univ. Press, (1997).
		\bibitem{Bernardini}A. E. Bernardini, S. S. Mizrahi, Phys. Scr. \textbf{90}, 074011 (2015).
	\bibitem{Sir}F. G. Scholtz, B. Chakraborty, J. Govaerts and S. Vaidya, \emph{J. Phys. A: Math. Theor.} \textbf{40} 14581 (2007).
\bibitem{fgs}F G Scholtz, L Gouba, A Hafver and C M Rohwer, \emph{J. Phys. A: Math. Theor.} \textbf{42}, 175303 (2009).
	
	\bibitem{Moyal} J. E. Moyal, 
	Proc. Camb. Phil. Soc. \textbf{45,} 99 (1949); H. Groenewold, 
	Physica \textbf{12,} 405, (1946).
	
	\bibitem{cris}L. C. B. Crispino, A. Higuchi, G. E. A. Matsas, \emph{Rev. Mod. Phys.} \textbf{80} 0034 (2008).
	\bibitem{partha}S. Biswas, P. Nandi, B. Chakraborty, Phys. Rev. A \textbf{102} 022231 (2020).
	\bibitem{smy}A. Hatzinikitas, I. Smyrnakis, J. Math. Phys. \textbf{43,} 113-125 (2002).
	\bibitem{Shahn}S. Majid, Class. Quan. Grav. \textbf{5,} 12, 1587, (1988); S. Majid,  J. Math. Phys. \textbf{41,} 06, 3892, (2000).
		\bibitem{niu}D. Xiao, J. Shi, Q. Niu, Phys. Rev. Lett. \textbf{95}, 137204 (2005).
	\bibitem{duvalberry}C. Duval, Z. Horvath, P. A. Horvathy, L. Martina, P. C. Stichel,	Mod. Phys. Lett. B \textbf{20}, 373-378 (2006).
		
\bibitem{fred}J.B. Geloun, S. Gangopadhyay, F. G. Scholtz,  Europhys. Lett. \textbf{86,} 51001 (2009).	

	
	
 
	

	\bibitem{Lvovsky}A. I. Lvovsky, arXiv:\textbf{1401.4118}, quant-ph (2014).
	\bibitem{coldatom}``Two-mode  squeezing in a cold atomic ensemble", University of Calgary Ph.D. thesis, written by A. Tashchilina (2019).
	
		

 
			\bibitem{baxter}C. Baxter, Phys. Rev. Lett. \textbf{74,} 4 0031 (1995).

	\bibitem{polychro}S. Ganeshan, A.P. Polychronakos, arXiv:\textbf{1912.10805v2,} hep-th (2019).
	
	\bibitem{Ban}M. Ban, J. Opt. Soc. Am. B, \textbf{10,} 1347 (1993).
	

	
	
	

	

	
	

	
\end{thebibliography}
\end{document}